\documentclass[figure]{pasj00}
\Received{$\langle$reception date$\rangle$}
\Accepted{$\langle$acception date$\rangle$}
\Published{$\langle$publication date$\rangle$}
\SetRunningHead{Polarimetric studies of comet Halley}{Polarimetric
studies of comet Halley}

\usepackage{times}

\begin{document}

\title{Polarimetric studies of comet Halley}
\author{Himadri Sekhar \textsc{Das} %
\thanks{Last update: March 31, 2007}}
\affil{%
   Department of Physics, Kokrajhar Govt. College\\
   Kokrajhar 783370, Assam, India}
   \email{hs\_das@rediffmail.com}
\and
\author{Sujit Ranjan \textsc{Das}%
}
\affil{%
   Department of Physics, Madurai Kamaraj University\\
   Madurai 625021, India}

\maketitle

\begin{abstract}
In the present work, the non-spherical grain characteristics of
comet Halley are analysed using the T-matrix method at $\lambda =
0.365, 0.485$ and $0.684\mu m$ respectively. In order to analyse
the polarisation data of comet Halley, the dust size distribution
function derived by Das et al. (2004) for comet Halley is used in
the present work. The size range of the grains is taken to be
0.01$\mu m \le$ s $\le 3\mu m$. Using the T-matrix method, the
best fit values of complex refractive indices $(n, k)$ and aspect
ratio ($E$) are determined at three different wavelengths $ 0.365,
0.485$ and $0.684\mu m$ and the corresponding values are given by
$(1.380, 0.043, 0.962), (1.378, 0.049, 0.962)$ and $(1.377,
0.058,0.962)$ respectively. After comparing the above result with
Mie theory result, it is found that prolate grains give the best
fit to the observed polarisation data of comet Halley. Also the
negative polarisation behaviour of comet Halley is discussed
thereafter.
\end{abstract}
\section{Introduction}
The study of polarisation of the scattered radiation from comets,
over various scattering angles and wavelengths, gives important
information about the nature of cometary dust grains. However at
certain wavelengths, the polarisation features are contaminated
due to the polarisation present in the cometary molecular line
emission. Since the last apparition of comet Halley, observers
have been using a set of filters (centered at $\lambda =0.365 \mu
m, 0.485 \mu m$ and $ 0.684 \mu m$ ) known as IHW (International
Halley Watch) filters to avoid contamination by line emission.

The analysis of polarisation data gives information about the
physical properties of the cometary grains, which include size
distribution, shape and complex refractive index. The \textit{in
situ} dust measurement of comet Halley gave the first direct
evidence of grain mass distribution (Mazets et al. 1986). Mukai et
al. (1987) and Sen et al. (1991a) analysed the polarisation data
of comet Halley using the power law dust distribution (Mazets et
al. 1986) and Mie theory, and derived a set of refractive indices
of cometary grains. The dust distribution function derived by
Mazets et al. (1986) is actually based on only Vega 2 results
while Lamy et al. (1987) derived the grain size distribution for
Halley by comparing the data from spacecrafts Vega 1, Vega 2 and
Giotto. However, Das et al. (2004) also analysed the polarisation
data of comet Halley using dust distribution function suggested by
Lamy et al. (1987).

The polarisation data of  several other comets were analysed by Das
et al. (2004) using Mie theory. They discussed the grain aging of
comets by solar radiation for four non-periodic comets (Hyakutake,
Austin, Bradfield, Levy 1990XX) and found out an empirical relation
between relative abundance of coarser grains index $(g)$ and
perihelion distance $(q)$ of the form $g = -2.5 q^{2/3}$. Das et al.
(2004) further commented that comets whose grains are processed more
by the solar radiation do contain relatively smaller number of finer
grains. From their work, it has been found that the grains of comet
Levy 1990 XX are much smaller, as compared to the grains of
Hyakutake, Austin, Bradfield, Hale-Bopp and Halley.

Several investigators made useful polarimetric measurements of comet
Halley through IHW filters (Bastien et al. 1986, Kikuchi et al.
1987, Le Borgne et al. 1987, Sen et al. 1991a, Chernova et al.
1993). The polarisation data of comet Halley were analysed using Mie
theory which assumes the dust particles to be spherical (Mukai et
al. 1987, Sen et al. 1991a, Das et al. 2004). But it is now accepted
that cometary grains are not spherical and may be \emph{fluffy
aggregates} or \emph{porous}, with irregular or spheroidal shapes
(Greenberg $\&$ Hage 1990). The measurement of circular polarisation
of comet Hale-Bopp (Rosenbush et al. 1997) also reveals that
cometary grains must be composed of non-spherical particles. Thus
the polarimetric data analysis using Mie theory will give less exact
results. In order to study the irregular grain characteristics of
comets, Discrete Dipole Approximation (DDA see viz. Draine 1988),
T-matrix theory (Waterman 1965) etc. are used. Xing $\&$ Hanner
(1997) have done elaborate calculations with porous grains of
different shapes and sizes using DDA method. Petrova et al. (2000)
have shown that aggregates composed of touching spheres with size
parameters 1.3 - 1.65 display properties typical of cometary
particles. Their results on the aggregates indicate that more
compact particles have a more pronounced negative branch of
polarisation. However, the DDA method requires considerable computer
time and memory. The T-matrix code on the other hand (Mishchenko et.
al., 2002) runs much faster and the results obtained can be tuned
easily since the input parameters to the code can be adjusted and
re­run in a short time. Using the T-matrix code, Kerola $\&$ Larson
(2001) analysed the polarization data of comet Hale-Bopp and found
the grains to be mostly prolate in shape in that comet. Recently Das
$\&$ Sen (2006) (our earlier work) using the T-matrix code found
that, the prolate grains can explain the observed polarization in a
better way as compared to the other shapes in comet Levy 1990XX.

In the present study,  the non-spherical grain characteristics of
comet Halley are analysed using Mishchenko's (1991, 1998) T-matrix
code. The results obtained from the T-matrix code are compared
with the Mie theory results. Also the negative polarisation
behaviour of comet Halley is discussed thereafter.


\section{Grain characteristics of comet Halley}
Polarimetry in the continuum has always been considered as an
important technique in the study of cometary dust properties. The
observed linear polarisation of comets is generally a function of
$(i)$ wavelength of incident light ($\lambda$), $(ii)$ Scattering
angle, $\theta$ (= $180^0 -$ Phase angle), $(iii)$ the geometrical
shape (E) and size ($s$) of the particle and $(iv)$ the
composition of dust particles in terms of complex values of
refractive index, $m$ $(= n - ik)$. The shape  of a spheroid can
be specified by the axial ratio, E ($= a/b$). It is to be noted
that  E $> 1$ for oblate spheroids, E $< 1$ for prolate spheroids
and E $= 1$ for spheres.
\subsection{In situ measurements of comet Halley}
During the last apparition of comet Halley, \emph{in situ}
analysis of comet Halley has been made possible. Several
spacecrafts on board Vega 1, Vega 2 and Giotto carried out
important measurements to determine the number density of
particles of given masses. Based on SP-2 experiment on-board Vega
space-craft, Mazets et al. (1986) had determined a set of power
laws ( with separate indices for different mass ranges ) for the
particle mass distribution over the range $10^{-16}$g to
$10^{-7}$g. Assuming the grain bulk density to be 1 g $ cm^{-3}$,
Mukai et al. (1987) derived the dust size distribution function
for comet Halley using the particle mass distribution suggested by
Mazets et al. (1986). Based on this size distribution and Mie
scattering formulation, Mukai et al. (1987) and Sen et al. (1991a)
analysed the polarisation data of comet Halley. However, Lamy et
al. (1987) combined the \emph{in situ} dust measurements from the
Vega 1, Vega 2 and Giotto and modelled the dust mass distribution
function.

Actually, the dust mass distribution function  suggested by Mazets
et al. (1986) is based on only Vega 2 results, while the work of
Lamy et al. (1987) is based on the results of three space-crafts.
Further, the size distribution function  derived by Mukai et al.
(1987) on the basis of the work reported by Mazets et al. (1986)
has three discrete size ranges and the size distribution function
changes its value abruptly over the three ranges due to the
presence of three distinct values of power law index (Das et al.
2004). But the size distribution function obtained by Lamy et al.
(1987) has a smooth behaviour. So, Das et al. (2004) followed the
work of Lamy et al. (1987) and derived the dust size distribution
function for comet Halley. Using that size distribution function,
they analysed the polarisation data of comet Halley and found out
a set of complex refractive indices $(n, k)$ which best match the
observed polarisation data.

The dust size distribution function $N(s)$ (with a bulk density of
dust grain, $\delta =$ 2.2 g $cm^{-3}$) for Halley derived by Das et
al. (2004), on the basis of work reported by Lamy et al. (1987) is
given by
\begin{equation}
\log N(s)= a (\log s)^2 + b (\log s) + c
\end{equation}

 where
$a=-0.2593$, $b=-4.422$ and $c=-15.06$.

It may be noted that  the grain size distributions used by Mukai
et al. (1987), or the one derived from Lamy et. al. (1987)(Ref.
Eqn 1) are basically the ones obtained from the last apparition of
comet Halley in 1985-86 and were used in explaining mostly the
polarisation properties of comets (Das et al. 2004).

Using Mie scattering theory and size distribution function (Eqn
1), Das et al. (2004) analysed the polarisation data of comet
Halley and determined the best fit values of $(n, k)$ at which the
sum of squares of difference between expected and observed values
of polarisation ($\chi ^2$-value) becomes minimum at $\lambda =
0.365, 0.485$ and $0.684\mu m$ respectively.

In the present work,  `spheroidal dust grain model' is proposed
for Halley to study the linear polarisation data of that comet
using the dust size distribution function (Eqn (1)) and the
T-matrix method.

\subsection{Spheroidal grain model}
It is now accepted that cometary grains are not spherical and may
be irregular or spheroidal in shapes (Greenberg $\&$ Hage 1990).
 As already discussed the T-matrix method provides a powerful tool to study the
 spheroidal grains in comets. This method was first introduced by Waterman
(1965) for studying electromagnetic scattering by single,
homogeneous non-spherical particles.  In this paper, calculation has
been carried out for randomly oriented spheroids using Mishchenko's
(1998) single scattering T-matrix code, which is available in
{\verb"http://www.giss.nasa.gov/"$\sim$ \verb"crmim". The important
feature of  T-matrix approach is that it reduces exactly to the Mie
theory when the particle is a homogeneous or layered sphere composed
of isotropic materials.

Several investigators studied irregular grain properties of comets
using T-matrix theory (Kolokolova et al. 1997, Kerola \& Larson
2001, Das \& Sen 2006). Kerola \& Larson (2001) analysed the
polarisation data of comet Hale-Bopp and found prolate grains to
be more satisfactory than other shapes in comet Hale-Bopp.
Recently, Das \& Sen (2006) studied the polarisation data of comet
Levy 1990XX using the T-matrix method and discovered that the
prolate shape of cometary grains can well fit the observed data.
Since no \emph{in situ} dust measurements were made on comet
Hale-Bopp and comet Levy 1990XX, power law distribution (Hansen \&
Travis, 1974) were used in both the cases. Further, the index of
refraction for olivine (1.63, 0.00003) was taken  in those two
comets for the analysis of polarisation data.

Using Mie scattering theory and grain model of Mazets et al.
(1986), Mukai et al. (1987) analysed comet Halley and found a set
of three complex refractive indices $(n, k)$ at three IHW filters
which best match their observation. Sen et al. (1991a) combined
their polarimetric observations with those of other investigators
and estimated $(n, k)$ values which are slightly different from
those  of Mukai et al. (1987). Based on the dust size distribution
function (eqn (1)) and Mie theory, Das et al. (2004) also analysed
the data and found a set of refractive indices $(n, k)$ for comet
Halley. Lamy et al. (1987) denoted the hypothetical refractive
indices $(n, k)$ emerging out from these Mie calculations as
`Silicate B'.

In the present work, the data is compiled on the polarisation
measurements of comet Halley that were made through IHW filters
and published in various journals (Bastien et al. 1986, Kikuchi et
al. 1987, Le Borgne et al. 1987, Sen et al. 1991a, Chernova et al.
1993). Here, equation (1) is considered for the grain size
distribution.

The detectors on-board the Vega and Giotto spacecrafts had
sensitivities as low as 10$^{-16}$g, and it was observed that the
particle number density continued to increase till the detection
limit was reached (Mazets et al. 1986). Assuming particles of
density 1 or 2.2 g cm$^{-3}$, one can find a lower limit for the
particle radius of $0.01 \mu m$ (Das et al. 2004).  Thus the minimum
radius of the grains can be fixed at $0.01\mu m$. It is to be noted
that in the present case, the T-matrix code can safely run on a
computer when the \emph{size parameter}, X (= $2\pi s / \lambda)$ is
less than 52. The choice of $\lambda = 0.365, 0.485$ and $0.684\mu
m$ gives the  maximum allowable radii of the dust grain to be
roughly 3$\mu m$, 4$\mu m$ and 5.5$\mu m$ respectively. In the
present work, the maximum  radius of the dust grains is thus fixed
at $3 \mu m$. Hence in order to analyse the polarimetric data of
comet Halley  at $\lambda = 0.365, 0.485$ and $0.684\mu m$
respectively, the size range of the grains is taken as $0.01 \mu m
\le s \le 3 \mu m$.

Using the T-matrix method, the best fit values of $(n, k)$ and E
are determined at which the sum of squares of differences between
calculated and observed values of polarisation ($\chi ^2$-value)
becomes minimum. These values are listed in \textbf{Table-1}. No
such good fit has been observed for oblate shapes. The
calculations are repeated for spherical grains (E=1), keeping $(n,
k)$ fixed, using Mie theory at $\lambda = 0.365, 0.485$ and
$0.684\mu m$ respectively. It is clear from \textbf{Table-1} that
data fits well for prolate grains with E=0.962 at $\lambda =
0.365, 0.485$ and $0.684\mu m$ respectively.

\begin{table*}

\begin{center}

 \caption{The $(n, k)$ values and E obtained in the present
work for comet Halley at different wavelengths.}
\begin{tabular}{|c|c|c|c|c|c|c|c|}
\hline
  $\lambda$   & Scattering angle& No. of data & $n$&$k$ & E& $\chi
  ^2_{min}$& Source of   \\
(in $\mu m$) &range (in degree)  &points &&&&& polarisation data \\
  \hline
  0.365& 114 - 178 &43&1.380&0.043&0.962&7.73& Bastien et al. (1986)\\
  \cline{6-7}
  & & & & &1.000&7.91& Kikuchi et al. (1987)\\
  \cline{1-7}
0.485&114 - 178&72&1.378&0.049&0.962&31.85& Le Borgne et al. (1987)\\
\cline{6-7}
 & & & & &1.000&32.25& Sen et al. (1991a)\\
  \cline{1-7}
  0.684&114 - 162&25&1.377&0.058&0.962&68.99& Chernova et al. (1993)\\
  \cline{6-7}
 & & & & &1.000&71.10&\\
  \hline
\end{tabular}
\end{center}
\end{table*}

Greenberg $\&$ Li (1996) studied interstellar dust polarisation
and found that prolate grains can give more satisfactory results
as compared to other shapes. Prolate spheroids are a natural
result of the process of \textit{clumping} in the proto-solar
nebulae (Kerola $\&$ Larson 2001). Thus the findings of prolate
grains in comet Halley strengthen the concept that cometary grains
are not of spherical shape.

In \textbf{Fig 1}, \textbf{Fig 2} and \textbf{Fig 3}, the expected
polarisation curves have been generated using the T-matrix code on
the observed polarisation values reported by various authors at
wavelengths 0.365, 0.485 and $0.684\mu m$ respectively.

\begin{figure}
\begin{center}
\FigureFile(80mm,50mm){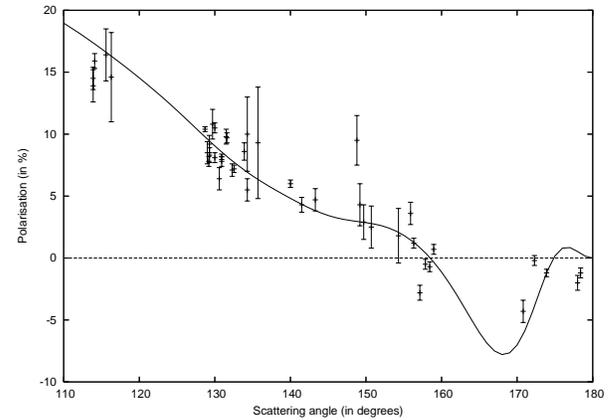}
\end{center}
 \caption{The observed polarisation values of comet Halley at $\lambda =
0.365 \mu m$. The solid line represent  the best fit polarisation
curve  obtained from the T-matrix code with $n$ = 1.380, $k$ =
0.043 and E = 0.962.
     }
\end{figure}

\begin{figure}
\begin{center}
\FigureFile(80mm,50mm){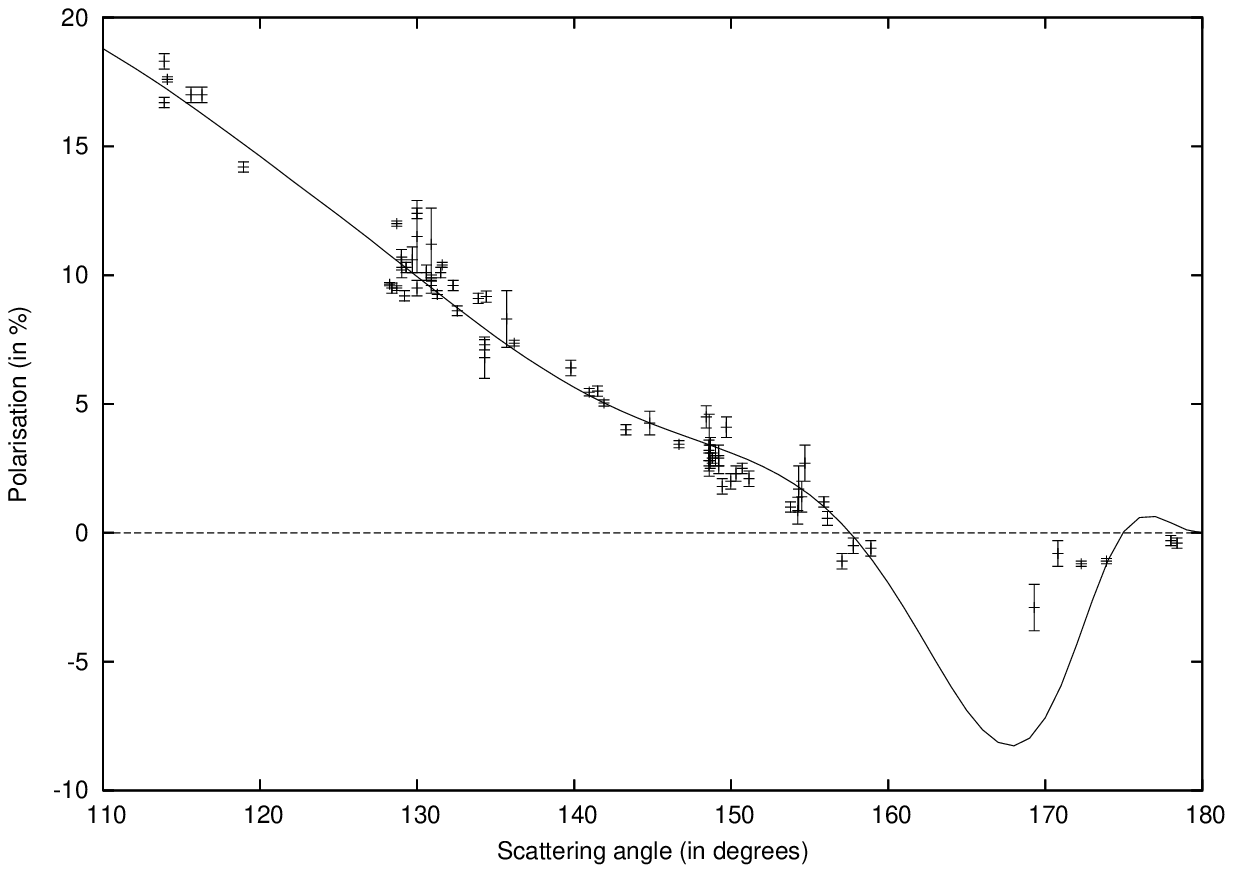}
\end{center}
 \caption{The observed polarisation values of comet Halley at $\lambda =
0.485 \mu m$. The solid line represent  the best fit polarisation
curve  obtained from the T-matrix code with $n$ = 1.378, $k$ =
0.049 and E = 0.962. }
\end{figure}

\begin{figure}
\begin{center}
\FigureFile(80mm,50mm){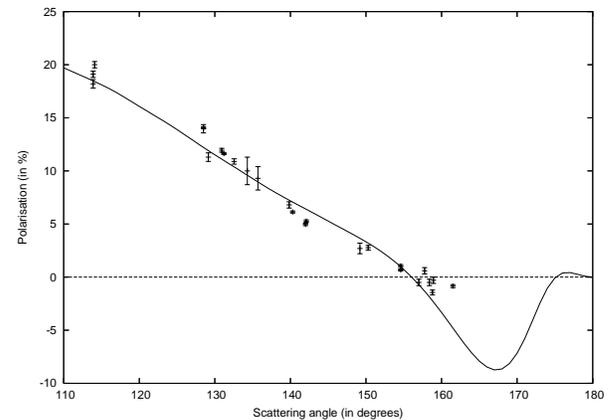}
\end{center}
 \caption{The observed polarisation values of comet Halley at $\lambda =
0.684 \mu m$. The solid line represent  the best fit polarisation
curve  obtained from the T-matrix code with $n$ = 1.377, $k$ =
0.058 and E = 0.962.  }
\end{figure}

\section{Discussions}
In the  present paper, a simple spheroidal model has been proposed
to study the non-spherical grain characteristics of comet Halley. It
can be seen from the present analysis that the prolate shape of
cometary grains are more satisfactory in comet Halley. This nature
of cometary grain is also observed in comet Hale-Bopp (Kerola \&
Larson 2001) and comet Levy 1990XX (Das \& Sen 2006). Since a good
number of polarisation data is available for comet Halley over a
wide scattering angle range, it is needless to say that the data
analysis will give more accurate results as compared to other
comets. To study other comets, Halley is  always taken as the
reference comet for all types of discussion. Thus, it can be
inferred from the above analysis that the dust grains in comets are
\emph{not perfectly spherical}.

The negative polarisation behaviour is one of the important
phenomena observed in comets. Several comets show negative
polarisation beyond the $157^0$ scattering angle (Kikuchi et al.,
1987; Chernova et al., 1993; Ganesh et al., 1998 etc.). Many
investigators (Greenberg \& Hage 1990; Muinonen 1993, Tanga et al.
1997, Levasseur-Regourd et al. 1998 etc.) have discussed the cause
of negative polarisation in comets. The \textit{coherent back
scattering} mechanism suggested by Muinonen(1993) has been used to
explain the negative polarisation. The \textit{fluffy aggregate
model} originally proposed by Greenberg and Hage (1990) and later
adopted by Xing and Hanner (1997) are also used for the study of
negative polarisation in comets. Tanga et al. (1997) and
Levasseur-Regourd et al. (1998) suggested that \emph{multiple
scattering} may well explain the negative polarisation because
lower polarisation is found in the near-nucleus region of comets
where dusty jets are most pronounced. Kerola \& Larson (2001) also
suggested that combination of viewing geometry effects and
enhanced multiple scattering might provide a quantitative
explanation of the negative polarisation beyond $160^0$. Many
investigators (Mukai et al. 1987; Sen et al. 1991a, 1991b; Joshi
et al. 1997; Das et al. 2004 etc.) have generated expected
polarisation curve using Mie theory that shows negative
polarisation beyond $157^0$. In the present work, the negative
polarisation values have been successfully generated in comet
Halley using the T-matrix code  for $\theta > 157^0$. Das \& Sen
(2006) analysed the comet Levy 1990XX using the T-matrix theory
which reproduced the negative branch of observed polarisation, but
their analysis using Mie theory did not show any negative
polarisation curve.

Greenberg \& Hage (1990) originally proposed the presence of large
numbers of \emph{porous} grains in the coma of comets to explain the
spectral emission at 3.4 $\mu m$ and 9.7 $\mu m$. Dollfus (1989)
discussed the results of laboratory experiments by microwave
simulation and laser scattering on various complex shapes with
different porosities. The results of \emph{in situ} measurements
carried out on the Giotto spacecraft at comet Halley (Fulle et al.
2000) and the analysis of the infrared spectra of comet Hale-Bopp
(Moreno et al. 2003) also agrees with the model of aggregates. It is
clear from recent modeling of optical (Kimura 2001, Petrova et al.
2000 etc.) and thermal-infrared observations (Lisse et al. 1998,
Harker et al. 2002), and especially from the Stardust returned
samples, that comet dust consists of irregular, mostly aggregated
particles.

Xing \& Hanner (1997) have done calculation with porous grains of
various shapes and sizes using DDA method. Moreno et al. (2003)
studied the composite grains using the DDA method  for modelling
comet Hale-Bopp's dust grains in the mid infrared spectrum. Gupta et
al.(2006) also studied the angular distribution of the scattered
intensity and linear polarization of composite cometary grains using
the DDA method. They used the size range `$s$' from 0.05 to 1.0 $\mu
m$, which corresponds to equivalent volume size parameter X = $2\pi
s / \lambda$ from 0.14 to 3.0 at the wavelength of 2.2 $\mu m$,
where `$s$' is the radius of the sphere of equivalent volume of the
host grain. However, the DDA method requires considerable computer
time and memory. The DDA code allows accurate calculations of
electromagnetic scattering from targets with size parameter X $< 15$
provided the refractive index $m$ is not large compared to unity
($|m-1| < 2)$ (Draine \& Flatau, 2004). It is to be noted that the
choice of $\lambda = 0.365 \mu m$ gives the value of `$s$' to be
less than $0.9 \mu m$. Thus using the DDA code, it is not possible
to study the scattering properties of composite grains if we
consider $s > 0.9 \mu m$. Using the N-sphere method, Petrova et al.
(2000) have shown that irregularly structured aggregates composed of
a moderate number of touching spheres ($< 50$) with size parameters
1.3 - 1.65 display properties typical of cometary particles. Their
results on the aggregates indicate that more compact particles have
a more pronounced negative branch of polarisation. The N-sphere
approach is based  on the calculation of clusters of T-matrices and
provides an accurate solution for randomly oriented arbitrarily
shaped aggregates of spherical monomers (Mackowski and Mishchenko
1996). The available computer restricted the size of the monomers to
a size parameter X = 2.5 only and the number of monomers within the
aggregate to 43. Thus, because of the need for a very large amount
of CPU time and memory storage, we could not extend our analysis to
published polarisation data of comet Halley at $\lambda = 0.365,
0.485$ and $0.684\mu m$ respectively using the DDA code and the
N-sphere approach, as the size parameter become too large to prevent
polarisation calculations with the available numerical codes and
computer facilities. The problems can be overcome only if a
significant improvement in the particle scattering codes and/or
computer speed takes place.

\section{Conclusions}
Based on the \emph{in situ} dust measurements and ground-based
polarimetric observations of comet Halley and also on the T-matrix
theory, the following conclusions can be drawn from the present
work:
   \begin{enumerate}
\item The complex refractive indices and shape parameter of
Halley's grains as derived from present work are: $(1.380, 0.043,
0.962), (1.378, 0.049, 0.962)$ and $(1.377, 0.058,0.962)$ at
$\lambda = 0.365, 0.485$ and $0.684\mu m$ respectively.

\item By observing $\chi ^2$-values in \textbf{Table-1}, one can
say that prolate grains can give  better fit to the observed
polarisation data.

\item The expected negative polarisation values have been
successfully generated for comet Halley using the T-matrix method.

\item The above model is suggested based on calculations which use
T-matrix theory, meaningful only for homogeneous spheroidal
particles. However, as cometary grains are \emph{porous}, a
follow-up paper is planned where calculations will be done with
more realistic \emph{porous} grains.

   \end{enumerate}

 \section{Acknowledgements}
 The author sincerely acknowledge IUCAA, Pune, where some of
 these calculations were done. The author is also thankful to M.
 Mishchenko for the T-matrix code.

\end{document}